\documentclass[manuscript, nonacm]{acmart}

\setcopyright{none}
\authorsaddresses{}

\AtBeginDocument{%
  }

\usepackage{booktabs}
\usepackage{tabularx}
\usepackage{enumitem}
\begin{document}

\title{The Doctor Will (Still) See You Now: On the Structural Limits of Agentic AI in Healthcare}
\titlenote{Preprint. Under review.}

\author{Gabriela Aránguiz Dias}
\authornote{Corresponding author: gadias@stanford.edu}
\affiliation{%
  \institution{Stanford University}
  \city{Stanford}
  \state{CA}
  \country{USA}
}

\author{Kiana Jafari}
\orcid{0000-0002-2001-3488}
\affiliation{%
  \institution{Stanford University}
  \city{Stanford}
  \state{CA}
  \country{USA}
}

\author{Allie Griffith}
\affiliation{%
  \institution{Stanford University}
  \city{Stanford}
  \state{CA}
  \country{USA}
}

\author{Carolina Aránguiz Dias}
\affiliation{%
  \institution{University of Florida}
  \city{Gainesville}
  \state{FL}
  \country{USA}
}

\author{Grace Ra Kim}
\affiliation{%
  \institution{Stanford University}
  \city{Stanford}
  \state{CA}
  \country{USA}
}
\author{Lana Saadeddin}
\affiliation{%
  \institution{Montclair State University}
  \city{Montclair}
  \state{NJ}
  \country{USA}
}

\author{Mykel J. Kochenderfer}
\orcid{0000-0002-7238-9663}
\affiliation{%
  \institution{Stanford University}
  \city{Stanford}
  \state{CA}
  \country{USA}
}

\renewcommand{\shortauthors}{Aránguiz Dias et al.}

 \begin{abstract}
Across healthcare, agentic artificial intelligence (AI) systems are increasingly promoted as capable of autonomous action, yet in practice they currently operate under near-total human oversight due to safety, regulatory, and liability constraints that make autonomous clinical reasoning infeasible in high-stakes environments. While market enthusiasm suggests a revolution in healthcare agents, the conceptual assumptions and accountability structures shaping these systems remain underexamined. We present a qualitative study based on interviews with 20 stakeholders, including developers, implementers, and end users. Our analysis identifies three mutually reinforcing tensions: conceptual fragmentation regarding the definition of `agentic'; an autonomy contradiction where commercial promises exceed operational reality; and an evaluation blind spot that prioritizes technical benchmarks over sociotechnical safety. We argue that agentic {AI} functions as a site of contested meaning-making where technical aspirations, commercial incentives, and clinical constraints intersect, carrying material consequences for patient safety and the distribution of blame.

\end{abstract}



\received{13 January 2026}

\maketitle

\section{Introduction}
Current industry discourse envisions a future of autonomous healthcare delivery propelled by agentic artificial intelligence (AI). Defined by goal-directed behavior and the capacity to execute multi-step tasks with minimal intervention, these systems are designed to plan, reason, and act independently, rather than merely responding to discrete prompts~\cite{ibm_what_2025, raptis_agentic_2025}. This technical shift has fueled a massive commercial surge: agentic AI startups raised \$3.8 billion in 2024, nearly tripling the previous year, and the the global market is projected to reach \$199 billion by 2034, with the healthcare segment alone is projected to reach \$5 billion by 2030~\cite{hertogh_agentic_2025, grandview_agentic_2025, definitive_healthcare_2025, menlo_ventures_2025}. Given that medical errors remain a leading cause of death in the United States \cite{makary2016medical}, decision support tools are increasingly being positioned as essential interventions to enhance diagnostic accuracy and patient safety. Promotional discourse positions these systems as solutions to longstanding healthcare challenges: reducing administrative burden, augmenting clinical decision-making, and enabling personalized care at scale~\cite{industry_what_2025,coran_2025}.

Beneath this speculatory framing, however, is a more complicated reality. Despite advances in technical performance on benchmark tasks, AI systems in healthcare often face a disconnect between impressive results in controlled research settings and limited adoption in clinical practice~\cite{sokol_artificial_2025,el_arab_bridging_2025}. This disconnect is compounded by weaknesses in current evaluation and regulatory practices. Most AI-related clinical trials remain single-center studies with small populations~\cite{ibrahim_reporting_2021}. Further, many AI-enabled medical devices enter the market through regulatory pathways that do not require prospective clinical evaluation, and recalls are concentrated among products lacking clinical validation~\cite{lee2025early}. As a result, benchmark performance fails to translate to real-world use.

The accountability implications of this evaluation gap remain underexplored. The characterization of tightly constrained, oversight-dependent systems as independent agents raises fundamental questions: who is responsible when marketed capabilities outpace operational reality, and how should evaluation frameworks account for this gap when the meaning of “agentic AI” itself remains contested across healthcare contexts?

Healthcare AI governance frameworks increasingly emphasize the importance of accountability structures, transparency mechanisms, and oversight processes~\cite{jacob_ai_2025,bodnari_scaling_2025,bartlett_towards_2024}. For instance, the National Academy of Medicine’s AI Code of Conduct emphasizes a clear delineation of responsibilities across the technology lifecycle~\cite{krishnan2025artificial}, while the American Medical Association’s pillars for responsible adoption focus on establishing formal executive accountability~\cite{lozovatsky_thomas_2025}. However, these initiatives implicitly assume a shared understanding of the systems being governed. When stakeholders do not agree on what `agentic' means or what behaviors warrant evaluation, accountability becomes difficult to assign and governance becomes difficult to operationalize. 

Prior work has examined AI evaluation in healthcare through technical~\cite{singhal_large_2023,nori_can_2023}, ethical~\cite{morley_ethics_2020}, and implementation lenses~\cite{wiens_no_2019,kim_establishing_2025}. While researchers have developed benchmarks for assessing large language model performance on clinical tasks~\cite{nejm_medagentbench_2025,arora_healthbench_2025} and proposed frameworks for trustworthy AI deployment~\cite{reddy_governance_2019,jacob_ai_2025}, less attention has been paid to how contested meanings shape accountability \textit{in practice}. This tension is exacerbated for systems positioned at the boundary between tool and agent, between assistance and autonomy.

In this paper, we present a qualitative study examining how agentic AI is defined, evaluated, and constrained across the healthcare AI domain. Our research draws on interviews with 20 stakeholders, including developers, physicians, and scholars. We analyze how `agency' is understood within different institutional contexts, and investigate consequences for accountability and oversight. Our analysis identifies three interrelated tensions:

\begin{itemize}[topsep=1pt, itemsep=1pt]
    \item  Stakeholders lack a shared understanding of what constitutes `agentic AI,' with interpretations varying by disciplinary position and organizational role. This instability allows developers, deployers, and regulators to locate responsibility elsewhere, obscuring who is accountable when systems fail.
    \item Systems described as `agentic’ in commercial discourse operate with sharply limited autonomy in practice.
    \item Prevailing benchmarks emphasize technical performance while failing to assess dimensions central to accountable deployment, including trust calibration, workflow integration, and risk management in safety-critical clinical contexts.
\end{itemize}

These tensions are mutually reinforcing. Instability in definition creates space for overclaiming, and both persist in part because evaluation frameworks do not measure what matters for accountability. We argue that agentic AI in healthcare functions as a site of contested meaning-making, where technical aspirations, commercial incentives, and clinical constraints intersect. Advancing meaningful accountability requires reconciling these divergent framings, developing evaluation practices that reflect institutional realities, and critically examining how the `agentic' label shapes expectations, liability, and the distribution of blame.

\section{Related Work}
Our work contributes to three intersecting bodies of work: (1) analysis of AI evaluation frameworks in healthcare, which document the misalignment between benchmark performance and clinical deployment; (2) scholarship on accountability and liability in AI-assisted decision-making; and (3) sociotechnical systems research that examines how technical abstractions obscure context-dependent harms. While the existing literature demonstrates that current evaluation methods of agentic AI do not entail real-world deployment success despite high benchmark performance, it lacks theory grounded in stakeholder perspectives and governance realities. This study seeks to understand how agentic AI is interpreted, deployed, and governed in clinical settings, providing insight into how these understandings should shape evaluation priorities in healthcare.
\subsection{Contested Meaning}
Industry definitions of agentic AI emphasize autonomy, goal-directed behavior, and multi-step task execution~\cite{ uipath_agentic_2025, ibm_what_2025} while academic frameworks foreground planning capabilities, tool use, and environmental perception~\cite{raptis_agentic_2025}. As \citet{selbst2019fairness} argue, ambiguous technical definitions become ``dangerously misguided when they enter the societal context that surrounds decision-making systems.'' Unclear boundaries delineating which systems constitute agents hinder communication, governance, and accountability when systems fail.

In healthcare, this ambiguity is compounded by the additional complexity of clinical work, where layered protocols, EHR infrastructures, and distributed clinical roles make system behavior difficult to localize or attribute \cite{patel2015cognitive}. Despite this, vendor marketing materials position agentic AI as capable of autonomous clinical reasoning~\cite{athenahealth_agentic_2025, kasshout_agentic_2024}. This disconnect can create accountability challenges that existing evaluation frameworks are not designed to address.

\subsection{Limits of Current Evaluation Frameworks in Healthcare}
Standard evaluation frameworks for medical AI prioritize technical performance metrics, which on their own inadequately capture how systems would perform in clinical environments~\cite{arora_healthbench_2025,meimandi2025measurement,hardy2025more}. Although traditional benchmarks such as MedQA~\cite{jin2021disease} showcase models' near-perfect accuracy on medical licensing examinations, researchers argue that this type of evaluation is an ``oversimplification of the real-world interaction between clinicians and patients''~\cite{nejm_medagentbench_2025,johri2025evaluation}. Focus on isolated accuracy creates a "de-contextualization" problem, as benchmarks ignore the infrastructural dependencies, such as EHR integration and layered clinical protocols, that dictate real-world system behavior. These frameworks fail to measure the "human-in-the-loop" labor required to catch system errors, leaving the failure modes of autonomous agency unmapped \cite{meimandi2025measurement}.

~\citet{arora_healthbench_2025} address this by introducing human-centered evaluation based on clinical judgment rather than factual correctness ~\cite{arora_healthbench_2025}. The authors find that high-performing models frequently overstate confidence, omit critical caveats, or fail to escalate to clinicians when there is uncertainty. MedHELM extends this work by evaluating across routine clinical tasks (documentation, workflows, patient communication), finding that performance varies considerably by task type; models excel at note generation but struggle with administrative workflows~\cite{bedi2025medhelm}. This variation attempts to address single-score claims of deployment readiness.

Beyond task selection, high accuracy does not guarantee reliable reasoning. Research on medical reasoning fidelity shows significant performance drops when familiar patterns are removed from benchmark questions~\cite{bedi2025fidelity}. \citet{kapoor2025holistic} demonstrate that agents with identical accuracy scores can exhibit vastly different risk profiles \cite{kapoor2025holistic}. Further, some agents complete tasks while violating instructions or taking unsafe intermediate actions, failures invisible to conventional metrics~\cite{kapoor2025holistic}. For instance, the authors document cases where a web agent benchmark assigns the same score of zero to both an agent that abstains from answering and one that leak's a user's credit card information during task execution, despite the vastly different consequences of these behaviors ~\cite{kapoor2025holistic}.

Research on multi-agent medical systems found that “best-of-breed” systems composed of individually strong-performing agents performed poorly when evaluated end-to-end ~\cite{bedi2025optimization}.Agents that rarely hallucinated in isolation produced hallucinations at rates approaching 14\% when integrated into pipelines, as upstream errors forced downstream components into fabrication \cite{bedi2025optimization}. 

A meta-analysis quantifies the measurement gap: 83\% of agentic AI evaluations focus on technical correctness, with far fewer assessing human-centered factors (30\%), safety and governance (53\%), or longitudinal behavior (5\%)~\cite{meimandi2025measurement}. The analysis documents cases where systems achieving 90–95\% accuracy in controlled testing showed limited adoption and negligible workflow impact post-deployment, echoing the \citet{selbst2019fairness} ``portability trap.’’

\subsection{Accountability Gaps}
Under current U.S. malpractice law, physicians bear responsibility whether they followed harmful AI recommendations or ignored helpful ones. This is a ``double bind’’ where clinicians are blamed regardless of how they engage with algorithmic advice~\cite{helzer_fault_2025,price_blackbox_2018}. Legal scholars propose that liability should depend on degree of autonomy~\cite{mezrich2022artificial,bottomley2023liability}, yet this framework presumes clear boundaries between \textit{assistive} and \textit{autonomous} systems. However, these boundaries are contested in practice.
\citet{habli2020artificial} argue that none of the actors in AI-assisted healthcare ``robustly fulfill the traditional conditions of moral accountability.’’ Governance frameworks respond by emphasizing clear accountability structures~\cite{duke_governance_2025}, yet these assume shared understanding of what is being governed. When stakeholders disagree on definitions, accountability becomes difficult to assign.

Prior work establishes that purely technical benchmarks fail to capture safety-critical dimensions, that agentic systems introduce invisible failure modes, and that accountability frameworks struggle to assign responsibility. However, less attention has been paid to how definitional ambiguity enables evaluation blind spots, and how both obscure accountability structures. We address this gap by grounding evaluation and accountability concerns in stakeholder perspectives, examining how meanings of agency are constructed across institutional contexts and what consequences follow for the allocation of responsibility.

\section{Methodology}
\label{sec:methodology}
\subsection{Research Design}
This study takes a grounded theory approach to investigate the development, use, and impact of agentic AI benchmarks in healthcare. Grounded theory allows conceptual insights to emerge inductively from participants’ perspectives rather than relying on predefined hypotheses ~\cite{khan2014qualitative, anderson2005empirical}. This approach was well suited to exploring how distinct stakeholder groups, ranging from biomedical data scientists to physicians, conceptualize and evaluate \textit{agentic} artificial intelligence systems in healthcare. We also relied on a semi-structured interview format, which allows for a balance between standardization across participants and flexibility to pursue unanticipated lines of discussion. This approach made it possible to develop themes that arise from both consistency across interviews and the spontaneous emergence of novel perspectives surrounding agentic AI in healthcare~\cite{adams2015conducting}.  

\subsection{Data Collection}
Participants were recruited from three main stakeholder groups involved in the development, evaluation, and use of healthcare AI systems: Technical Developers, Clinical Implementers, and Healthcare End Users. These categories encompass the end-to-end pipeline within healthcare contexts:

\begin{itemize}[topsep=0pt, itemsep=0pt]
    \item \textbf{Technical Developers} include biomedical engineers, computer scientists, and machine learning researchers who design, prototype, or evaluate agentic AI systems. Their perspectives inform how agency is technically instantiated, constrained, and measured within current development pipelines.
    \item \textbf{Clinical Implementers} include physician–developers, clinical data scientists, and translational researchers who integrate AI systems into existing healthcare workflows. They provide insights into how institutional, regulatory, and infrastructural factors shape the deployment and oversight of agentic AI.
    \item \textbf{Healthcare End Users} represent clinicians, medical professionals, and physician-scientists who interact directly with AI tools in diagnostic, operational, or treatment contexts. Their experiences reflect how AI autonomy is perceived, negotiated, and trusted in everyday clinical practice.
\end{itemize}
\begin{table}[htbp]
\small
\caption{Demographics of Interview Participants ($N=20$)}
\label{tab:participant_demographics}
\renewcommand{\arraystretch}{0.9} 
\begin{tabular}{@{} l l l l @{}}
\toprule
\textbf{Code} & \textbf{Role} & \textbf{Gender} \\ \midrule
I-1 & AI/ML Research Director, Federal Health IT & Male\\
I-2 & Medical Student, Clinical End User & Male\\
I-3 & AI Researcher & Male\\
I-4 & Healthcare AI Professional & Male\\
I-5 & Physician, AI Researcher & Male\\
I-6 & MD-PhD Student, Biomedical Data Science & Male\\
I-7 & CEO/Co-founder, Healthcare AI Startup& Male\\
I-8 & Clinical Informatics Fellow & Male\\
I-9 & PhD Student, Biomedical Data Science & Female \\
I-10 & PhD Student, Digital Health/Behavioral Sciences & Male\\
I-11 & AI Fellow, Ethics and Policy & Male\\
I-12 & Postdoctoral Researcher, Medical AI & Male \\
I-13 & AI Strategy Consultant & Male\\
I-14 & AI Safety Researcher & Male\\
I-15 & AI Executive, Enterprise AI Strategy & Female\\
I-16 & Professor of Medicine, Chief Data Scientist & Male\\
I-17 & PhD Student/Researcher, Clinical AI & Female\\
I-18 & Nurse Practitioner, Clinical End User & Female\\
I-19 & Professor, Biomedical Engineering/Neuroscience & Male \\
I-20 & Computational Medical Imaging Research Scholar
 & Male\\ \bottomrule
\end{tabular}
\end{table}
\vspace{-0.8em}
In total, twenty participants were interviewed. 
This sample size was selected based on standard best practices for qualitative, interview-based studies in human-computer interaction literature \cite{jo2023understanding, muller2019data}. Interviews lasted approximately 45–60 minutes and were conducted by video conference between August and December 2025. All interviews were conducted in English, and all interviewees were based in the United States. All participants had firsthand experience with AI systems in healthcare, in addition to baseline familiarity with agentic AI systems and how they are being used in healthcare broadly. 

Participants were identified through a combination of both intentional and snowball sampling. Intentional sampling provided coverage of the three stakeholder groups above, while snowball sampling allowed recruitment to expand organically through interviewee referrals. This approach facilitated the inclusion of participants with distinct priorities that occupied diverse institutional roles. Recruitment continued iteratively until the authors determined thematic saturation was achieved; i.e, when no substantively new categories emerged from additional interviews. 

All participants provided informed consent, and participant identities were anonymized in the final dataset to protect their privacy. IRB approval was obtained prior to data collection, ensuring the study adhered to ethical guidelines for research with human participants (IRB-80503).

\subsection{Interview Design}
Interviews followed a semi-structured format organized around five thematic areas designed to capture both conceptual and experiential perspectives on agentic AI in healthcare: (1) framing and definitions of agency; (2) technical capability and perceived autonomy; (3) practical use and experience with deployed systems; (4) trust, oversight, and human–AI relationships; and (5) gaps, challenges, and future needs.  All participants were asked foundational questions about definitions, trust, and future directions, while role-specific sections were emphasized according to participant expertise, technical capability questions for developers and researchers, and practical experience questions for clinical end users. This role-aware structure allowed for the extraction of domain-relevant insights while preserving comparability across the dataset.

Interviews began by eliciting participants’ own definitions of agency and experiences with AI systems, grounding subsequent discussion in professional context. Questions then probed the extent of autonomous operation in current systems, factors shaping trust and oversight decisions, and perceived gaps between system capabilities and deployment realities. The semi-structured design allowed flexibility to pursue emergent themes while maintaining analytic consistency across the dataset. Each interview was conducted by one to two researchers, lasted 45–60 minutes, and was audio-recorded and transcribed verbatim with participant consent. The complete interview protocol is provided in Table~\ref{tab:interview_protocol_final} (Appendix).

\subsection{Data Analysis}
We analyzed interview transcripts using grounded theory coding. Transcription yielded a corpus of approximately 577 single-spaced pages (173,009 words), with individual interviews ranging from 4,691 to 13,255 words. Our analysis began with open coding to identify salient themes and concepts, followed by axial coding to establish relationships between emerging categories. During the analysis, we extracted passages containing substantive claims, evaluation practices, deployment experiences, or trust dynamics. Through the coding process, we identified connections between definitional fragmentation and evaluation gaps, between trust concerns and oversight requirements, and between institutional constraints and deployment barriers.

The final coding scheme comprised 6 high-level categories and 67 codes (see section~\ref{sec:coding_framework} for the complete codebook). Major categories included: (1) \textbf{agentic definitions and AI use}, capturing how participants conceptualized agency and their experiences with AI systems in healthcare; (2) \textbf{barriers and evaluation}, documenting technical bottlenecks, adoption challenges, evaluation metrics, and oversight mechanisms; (3) \textbf{trust and independence}, addressing factors influencing clinician trust, transparency requirements, and comfort with autonomous action; (4) \textbf{concerns}, cataloging safety-critical worries, institutional constraints, and integration challenges; (5) \textbf{expectation gaps}, capturing misalignments between capabilities and user expectations; and (6) \textbf{future directions}, documenting participants' visions for agentic AI development.

The analysis followed an iterative process, with the research team continually revisiting and revising codes based on emerging data. A thematic analysis approach ensured that the study captured the richness of interview data while building grounded theory reflecting participants' experiences. Two coders independently coded a subset of transcripts, with coding discrepancies resolved through team discussions. We used Taguette for qualitative coding and thematic analysis.

\section{Findings}
Analysis of interviews reveal that the evaluation of agentic AI in healthcare is misaligned with deployment realities due to persistent definitional ambiguity, structural constraints on autonomy, and evaluation practices that fail to capture trust, workflow integration, and domain-inherent error intolerance. Participants describe substantial ambiguity around what constitutes agency in healthcare AI, alongside near-universal constraints on autonomy compelled by a suite of safety, regulatory, and liability considerations. In the sections that follow, we present how these conceptual, structural, and evaluative gaps shape the assessment and adoption of agentic AI systems in healthcare settings.
\subsection{Conceptual Foundations \& Misalignment}
In this section, we present how agency in AI systems are currently conceptualized in healthcare. We found a striking absence of a shared definition of what `agentic AI' means in healthcare, with interpretations shaped by disciplinary background, institutional role, and practical constraints of clinical work. This conceptual fragmentation underlies downstream tensions around autonomy, evaluation, and trust.

\subsubsection{The Definitional Crisis: What is `Agentic AI?'}
Across interviews, participants expressed substantial disagreement on what constitutes `agentic AI’ in healthcare. Several characterized the term as a moving target or marketing term used to describe capabilities that current systems do not yet possess. Participant I-16 positions agentic systems as ``kind of a buzzword,’’ noting that limitations are repeatedly deferred into future framings of agency:
\begin{quote}
\vspace{-0.2cm}
    In 2024 whatever LLMs could not do, we said agents will do in 2025, and what agents cannot do in 2025 we say agentic systems will do in 2026.
\vspace{-0.2cm}
\end{quote}
Other participants highlight that disagreements around agency stem from distinct definitions across domains. Technical Developers define agents as entities with planning, problem decomposition, memory, tool use, and retrieval capabilities. Participant I-14  emphasizes the decomposition of tasks and tool use components:
\begin{quote}
\vspace{-0.2cm}
    [an agentic AI system is] an AI system that has some form of... tool access or embodiment, to then automatically do the task by coming up with a relevant subtasks, scheduling them, and then doing them in a consecutive order.
\vspace{-0.2cm}
\end{quote}
On the other hand, Healthcare End Users and Clinical Implementers describe agency in far more constrained terms. Participant I-9 delineates between these perspectives, explaining that healthcare definitions of agentic AI tend to focus on ``automating a workflow with concrete steps'' rather than open-ended autonomous reasoning. In this sense, agency is deliberately limited, or even ruled out entirely, to ensure predictability, safety, and alignment with clinical processes. 

Several participants attempt to reconcile these divergent interpretations by framing agency as a spectrum rather than a binary property. From this perspective, systems may exhibit varying degrees or types of agency. As participant I-7 frames it, these degrees of agency can range from “one-directional” agency, such as passive listening or documentation tools, to “bi-directional” agency involving feedback loops, and further to conversational agents capable of more interactive behavior. These spectrum-based accounts reflected efforts to accommodate heterogeneous systems under a single conceptual umbrella, though participants differed in where they placed boundaries between categories.

Other interviewees reject the label `agentic' altogether when applied to most current healthcare AI systems. From this view, agentic labels imply the ability to coordinate multiple tasks toward a specific outcome, a capability that participants argued is largely absent in deployed systems. Notably, participant I-11 asserts: `` I would not even call them agentic... an agentic system implies doing multiple tasks, chaining together for an outcome, and most systems are only good at one task.'' This perspective stresses that the vast majority of healthcare AI tools remain narrowly scoped to individual functions rather than exhibiting integrated, multi-step behavior.

Even within healthcare, the term `agentic' is applied inconsistently. Participant I-5 notes that the definition has been ``overused'' and has also ``been used in a couple of different ways.'' They go on to describe how the term can be alternately applied to systems that can “act in the world” and to collections of AI algorithms that ``work together as a team with different areas of specialization.'' This multiplicity of meanings further complicates attempts to evaluate or compare agentic AI systems, as different stakeholders may implicitly refer to distinct capabilities with the same terminology.

This definitional instability poses a significant challenge for evaluation and accountability. When stakeholders lack a common understanding of what behaviors, risks, or capabilities should be assessed, the same system can be simultaneously characterized as `agentic’ or `not agentic’ depending on who is asked.

\subsubsection{The Autonomy Contradiction: Agency Without Autonomy}
Despite widespread use of the term `agentic’ in healthcare marketing and industry discourse~\cite{athenahealth_agentic_2025, awsAgentic, geHealthcareAgentic}, we found that most deployed systems operate as assistants with firm human-in-the-loop requirements. `Agentic’ often signals a design aspiration or product positioning rather than operational reality. This disconnect between how agency is invoked and how systems are actually used presents a contradiction: AI systems are being presented as increasingly agentic, but they remain  constrained by safety, liability, and workflow requirements that necessitate human oversight.

We find that fully autonomous AI systems do not exist in clinical healthcare settings. Nearly every participant confirmed that current systems require ongoing human oversight, particularly for diagnostic and high-stakes clinical decision-making. Participant I-17 characterized current systems as ``0\% autonomous,’’ adding that ``there's always a physician in the loop.’’ Participant I-11 similarly noted that ``all of them require human oversight... I don't know of any that will make a clinical judgment by itself.'' Participants contextualized this reliance on oversight beyond technical limitations. As I-7 explained:
\begin{quote}
    If there isn't a human in the loop, … society does not have an appropriate framework for letting AI, letting something that's sometimes wrong, operate in a high-stakes environment.
\end{quote}
Even in lower-risk use cases, such as administrative tasks or documentation, autonomy remains tightly constrained. Inaccuracies in documentation may not seem immediately clinically harmful but can propagate into serious downstream consequences over time. Participant I-7 warned that using AI to document patient history ``without any human oversight’’ and immediately insert it into permanent records would result in ``terrible mistakes... with a large enough N, right? Such that lives would be severely impacted." I-7 reiterated that ``you cannot flip from a reactive system to a proactive system... without human oversight in areas of high risk.’’

Participants emphasizes that the field is not currently equipped for autonomous agents. Participant I-8 sharing that ``We're barely scratching the surface... we're not very advanced as a field in the ability to automate workflows in clinical care,'' while I-20 states frankly, ``we don't like autonomy here.'' This constraint is also evident in formal approvals; participant I-10 noted that only one LLM-based system has been approved as a medical device in the European Union is a retrieval-based question-answering system operating under direct clinician oversight.

The gap between marketing and deployment reflects a deeper misalignment. As participant I-11 observed:
\begin{quote}
    ...a lot of the people who are building for healthcare don't understand how healthcare actually works... oftentimes, inside looking out, they don't really need the [solution] you're trying to build.
\end{quote}
Agentic AI in healthcare is less about independent action and more about tightly bounded assistance, with interpretation, judgment, and intervention left to humans. 
\subsection{Evaluation, Trust, and Safety Constraints}

Across interviews, evaluation, trust, and safety emerged as inseparable themes shaping whether agentic AI can be meaningfully deployed in healthcare. Existing evaluation approaches, largely inherited from machine learning benchmarks, are widely seen as insufficient for capturing the reliability, oversight, and contextual specificity required in clinical settings~\cite{meimandi2025measurement}. These evaluation gaps contribute directly to lack of trust, which in turn constrains the autonomy systems can safely assume.
\subsubsection{The Evaluation Gap: ``We Don't Know What Metrics Matter''}
Participants emphasized that the field lacks clarity on what should be evaluated in the first place. Participant I-16 articulates this uncertainty, noting that ``we're still at a stage where we don't know what are the metrics that matter,'' and questioning whether evaluation should focus on individual agent behaviors or system-level performance.
Participant I-17 describes evaluation practices as ``very vague,'' emphasizing that clinical judgment varies depending on specialty, context, and risk tolerance, inhibiting the collection of ground-truth labels.
Participants further highlight that existing evaluation suites fail to capture the realities of clinical deployment. 
Participant I-3 points to practical constraints noting that ``even if performance looks okay technically, issues like latency and workflow integration become barriers.'' Participant I-9 underscores that 
``verification is probably a bigger bottleneck than generation.'' The consequences are severe for diagnostic applications. Participant I-20 hones in on the ecological validity of relying on publicly available benchmarks to make diagnostic determinations: ``\dots those are, hand-picked. The outliers are excluded. And we trained on those datasets, and then if you put it into a real [practice], yeah, it's a disaster.''

Structural challenges compound these limitations as system complexity increases. Participant I-5 emphasizes that meaningful evaluation becomes especially difficult when human oversight is reduced or multiple agents interact:
\begin{quote}
\vspace{-0.2cm}
    [With] no humans in that loop, we're gonna\dots need some serious evaluation, and that work has not been done yet. And that work is, by the way, extremely difficult when you have multiple agents all working together. How do you evaluate that system?
\vspace{-0.2cm}
\end{quote}
In the absence of standardized metrics, evaluation defaults to pragmatic human-centered proxies. Participant I-7 describes relying on ``dedicated physicians, where whenever we have something new that we're gonna change and release, we communicate it with that tight user group, and they'll test it first\dots and then they give focused feedback.’’ Human judgment thus becomes both evaluative mechanism and safety backstop; a dynamic that limits scalability and constrains autonomy.

We find that gaps in evaluation and metrics contribute to an overall lack of trust, despite trust being essential to the adoption and deployment of automated systems in healthcare.

\subsubsection{Trust as the Primary Barrier}
Where formal metrics fall short, \textit{trust} and \textit{trustworthiness} emerge as dominant evaluation criteria.  Yet trust itself is rarely defined, measured, or operationalized. Participant I-5 articulates the critical distinction: ``there's a difference between trust and trustworthiness. So, you can make a user trust something, but it doesn't necessarily deserve the trust.''

For healthcare end users, systems that appear trustworthy by technical metrics often fail to earn trust in practice. Trust in AI systems, particularly those with increased autonomy, is shaped by clarity around liability and responsibility, interpretability of system decisions, reliability over time, alignment with clinical workflows, and transparency around corporate and profit motives. 
Participant I-19 explains that ``the surgeon may not trust the tool\dots at the end of the day, they are on the hook.’’ Because clinicians remain legally and ethically responsible for patient outcomes regardless of whether decisions are AI-assisted, trust in AI systems is necessarily conservative.

Reliability expectations create particular tension. Participant I-7 explained that clinicians ``want determinism,’’ and ``the fact that [the system] is wrong sometimes\dots imposes a ceiling\dots I have no idea how that ceiling could be replaced.’’ This stands in tension with the probabilistic nature of any AI-powered system. 

Trust depends not only on whether a system produces correct outputs, but on whether clinicians can understand why a particular recommendation or action was produced. Participant I-5 adds that clinicians want visibility into ``the chain of reasoning,’’ especially in high-stakes decisions where accountability cannot be deferred to black-box models. Without this kind of transparency, clinicians lack the means to adequately understand or contest system behavior when outcomes are uncertain.

Trust is further shaped by concerns about system integrity and control, particularly in agentic settings where systems may act across multiple steps or data sources. Participant I-13 raises concerns about data corruption and ``rogue strategies,’’ reflecting anxiety that autonomous systems could behave unexpectedly outside narrow constraints. 

There is also skepticism rooted in institutional and economic considerations. Participant I-2 questions ``the reliability of private companies to interact in personal networks in a way that is separate from the profit motive,’’ while I-14 predicted that ``I would put on the cynic hat and assume the key main driver will be economic incentives.’’

All of these trust concerns culminate in a broader affective response among clinicians: fear, skepticism, and a redefinition of their role in relation to the agentic AI systems on the horizon. 
Participant I-9 captures how fear and role uncertainty shape engagement:
\begin{quote}
\vspace{-0.2cm}
    We're scared. There's skepticism of, oh, is it going to replace doctors? Oh, is it going to be a companion? Is it going to be augmentative?\dots people are now reaching the consensus [that] these tools are just our, like, companions or assistants,\dots they're not gonna replace us. But they're actually taking up part of a clinician's role they used to do before, and their role is evolving more into\dots chaperoning, or\dots making sure that the system is not brittle. So I think right now, it's going to be a little bit more reactive, because we don't trust them.
\vspace{-0.2cm}
\end{quote}
\subsubsection{The Hallucination Problem and Error Tolerance}
Healthcare’s low tolerance for error makes hallucinations and instability particularly problematic. Participant I-10 highlights that ``Hallucinations are a problem even with temperature at zero\dots omission of information is equally important\dots if you're missing the milligram amount of a medication, that is immediately dangerous.''

Beyond isolated errors, many participants raise concerns about inconsistency and instability in agentic and multi-agent systems. Participant I-16 describes a stark lack of consistency across multi-agent experiments: "You ask the same question about the same patient, what trials are they eligible for? Sometimes it says four trials, sometimes zero.'' 

This kind of variability erodes trust in settings where reproducibility is critical for safe decision-making. Some participants cited hallucinations as a reason to forego any kind of agency altogether. Rather than allowing systems to act independently, Participant I-1 advocates for ``AI in the loop,'' where AI systems monitor, summarize, or surface information while humans retain full decision authority. 

Overall, we observe that as long as hallucinations and instability remain unresolved, increased agency is viewed not as a benefit, but as a serious risk. 

\subsection{Operational Constraints and Viable Use Cases of Agentic AI}
Beyond ambiguity in definition and evaluation gaps, participants identified operational constraints that determine which use cases are viable today and which remain aspirational.
\subsubsection{The Workflow Integration Challenge}\label{sec:workflow_integration}
Systems often fail not because they lack capability but because they cannot integrate into existing clinical workflows. Participant I-9 illustrated this with a simple example:
\begin{quote}
\vspace{-0.2cm}
An example is, I want to order a wheelchair for a patient\dots just that task involves multiple entities\dots nurses, doctors, administrators interacting with each other. There are a lot of forms that you have to hand off, and that whole process can take a while.
\vspace{-0.2cm}
\end{quote}
Wheelchair ordering involves insurance verification, equipment availability, discharge planning, and documentation across multiple systems. An agent that automates only the form-filling misses the relational work required.  Infrastructure limitations compound these challenges. Participant I-8 emphasizes that healthcare data remains fragmented:
\begin{quote}
\vspace{-0.2cm}
We are just scratching the surface of what's needed in those spaces for an agent to successfully be able to traverse multimodal data in clinical care\dots We're not very advanced as a field in the ability to automate workflows in the clinical space.
\vspace{-0.1cm}
\end{quote}
Participant I-15 adds that vendor-built agentic systems perform well only within their native platforms but clinical workflows rarely confine themselves to single platforms:
\begin{quote}
\vspace{-0.2cm}
We really believe that all these agentic systems built by these vendors should be used for what they were built and close to what the data is\dots But if you want to think about [them] as a generic platform that can reason about the world, I think that is the wrong sort of use.
\vspace{-0.2cm}
\end{quote}
\subsubsection{Where Agentic AI Actually Works}
Despite these constraints, participants identified domains where agentic AI has demonstrated viability. These domains share common characteristics of lower stakes, well-defined workflows, and preserved human oversight.

\textit{Documentation and Administrative Tasks.} The most successful applications involve clinical documentation (ambient AI scribes) and administrative automation (prior authorization, billing, scheduling). Participant I-9 observes that startups concentrate here precisely because stakes are lower:
\begin{quote}
\vspace{-0.2cm}
A lot of startups are focusing on agents for prior authorization, for billing\dots These do not generally deal with patient safety\dots They're taking this approach of following an application that doesn't have too much of a harm if things go wrong.
\vspace{-0.2cm}
\end{quote}
This risk-based deployment strategy, starting with low-stakes tasks before advancing to clinical applications, emerged as a consensus approach.

\textit{Clinical Decision Support.} Systems that surface relevant information or suggest differential diagnoses have found adoption, but always with explicit human approval. Participant I-19 described a deep brain stimulation programming assistant that cuts clinician time from fifteen to five minutes per patient but ``the clinician still reviews it'' before changes take effect.

\textit{What Remains Out of Reach.} Participants were equally clear about domains where agentic AI has not succeeded: autonomous diagnosis, complex multi-step clinical reasoning, and real-time intervention in acute settings. As I-9 puts it, ``I don't think we're ever going to want to automate diagnosis, or have agents start from diagnosing a patient to prescribing a medicine.''

These boundaries reflect accountability structures and liability frameworks as much as technical limitations. Until those structures evolve, agentic AI in healthcare will remain confined to augmentation rather than automation.

\subsection{Temporal Dimensions of Evaluation}
A recurring theme across interviews concerns how evaluation frameworks fail to account for the temporal structure of clinical tasks. Healthcare AI systems must operate across vastly different timescales, from instantaneous surgical decisions to longitudinal patient management spanning years; yet current evaluation practices remain largely static.

\subsubsection{Timescale Variability}
Participants identified a fundamental mismatch between the temporal demands of clinical work and current capabilities. Participant I-19 articulated this gap:
\begin{quote}
In agentic AI systems you might run into things where you have a long timescale of decision making process. You're processing patient records from since they started the admission process\dots all the way until, maybe could span tens of years\dots In terms of short time scales, you have to be making instantaneous decisions that take the observation of the environment at the current moment to make very well informed decisions.
\end{quote}
A system optimized for rapid triage faces fundamentally different requirements than one designed to synthesize a patient's multi-year treatment history. I-19 observes, ``most of the current agent AIs are lacking all of this, conscious awareness of the importance of timescale when making those decisions.'' I-19 goes on to argue that while time is "one dimensional" and "not very complicated," it remains a critical yet neglected variable in agentic decision-making and the development of agentic systems in healthcare.

This neglect has practical consequences. The temporal sequence of observations often carries as much diagnostic information as the observations themselves. Participant I-16 highlights how temporal context shapes acceptable performance: a system with moderate error rates might be tolerable in outpatient primary care, where follow-up allows course correction, but unacceptable in the emergency room, where decisions are often irreversible.

\subsubsection{The Static Benchmark Problem}
Participants criticize current evaluation practices as overly static. I-14 identified this as a fundamental limitation, advocating for a shift from ``tiny corner[s] of scenario[s]'' toward ``longer interactions'' and ``more dynamic settings'' that better reflect real-world clinical behavior. Static benchmarks such as single-turn question-answering tasks and fixed case vignettes cannot assess how systems behave over extended interactions or adapt as new information arrives. I-19 connects this to the challenge of learning from sparse clinical data: ``
We need new algorithms that can learn from very limited amounts of data and be able to predict reliably\dots a fixed model... will [not] be able to perform reliably over the long term.''
These findings suggest that meaningful evaluation of agentic AI in healthcare requires explicit attention to temporal structure. This includes distinguishing between task categories operating on different timescales, assessing temporal consistency as patient data accumulates, and weighting risk assessment according to how long errors persist in the clinical record.

\section{Discussion}
Our findings reveal that definitional instability enables overclaiming, overclaiming persists because evaluation frameworks fail to measure what matters, and both dynamics obscure accountability when systems fail. When developers define agency in terms of planning and multi-step reasoning, while clinical implementers define it as bounded workflow automation, the same system can be simultaneously `agentic' and `not agentic,' depending on who is asked. This ambiguity has material consequences for responsibility allocation.
\subsection{The Accountability Vacuum}
Our findings extend the observation of \citet{habli2020artificial} that no actor in AI-assisted healthcare robustly fulfills the conditions of moral accountability. We show that definitional instability compounds this problem by allowing each stakeholder group to locate responsibility elsewhere. Developers claim their systems are merely tools; deployers point to vendor marketing promising autonomous capabilities; clinicians bear liability regardless of how they engage with algorithmic advice. The result is \textit{distributed lack of accountability}, responsibility diffused across so many actors that it effectively belongs to no one.

This carries implications for emerging governance frameworks. The National Academy of Medicine's AI Code of Conduct~\cite{krishnan2025artificial} and AMA accountability principles~\cite{lozovatsky_thomas_2025} assume stakeholders share an understanding of what is being governed. Our data suggest this assumption is unfounded. Meaningful accountability requires first establishing definitional consensus, or at minimum, requiring explicit disclosure of what capabilities a system possesses versus what marketing materials claim.
\subsection{The Autonomy Paradox}
Our findings document a striking paradox: systems described as increasingly agentic in commercial discourse operate with decreasing autonomy in deployment. Participants consistently reported that deployed systems function as tightly constrained assistants requiring human oversight, even as vendor materials position these same systems as capable of autonomous clinical reasoning.

This gap raises questions about promissory governance, whereby systems are governed based on anticipated future capabilities rather than current operational reality. When evaluation frameworks respond to promissory claims rather than deployed behavior, accountability becomes untethered from actual performance.

The autonomy paradox also illuminates why the translational gap persists despite high benchmark performance~\cite{el_arab_bridging_2025,sokol_artificial_2025}. Systems succeed on benchmarks measuring capabilities they will never be permitted to exercise. A system achieving 95\% accuracy on autonomous diagnostic tasks provides little information about deployment readiness when it will always operate under direct clinician oversight. Evaluation frameworks must assess systems as they will be used, not as promotional materials suggest.
\subsection{Toward Deployment-Centered Evaluation}
Participants emphasized that prevailing benchmarks fail to capture dimensions central to accountable deployment such as trust calibration, workflow integration, temporal consistency, and risk management. The measurement imbalance documented by Meimandi et al.~\cite{meimandi2025measurement}, wherein 83\% of evaluations focus on technical correctness while only 5\% assess longitudinal behavior, reflects deeper assumptions about what matters for deployment readiness.

We propose a shift from \textit{capability-centered} to \textit{deployment-centered} assessment. First, evaluation must account for sociotechnical context. Accountability cannot be abstracted from institutional settings~\cite{selbst2019fairness}. Our wheelchair-ordering example~(see section \ref{sec:workflow_integration}) illustrates how technically capable agents fail when they cannot navigate the relational work and cross-system coordination that clinical workflows require.

Second, evaluation must distinguish between trustworthiness and trust. Factors shaping trust, clarity around liability, interpretability of reasoning, demonstrated reliability, transparency about commercial motives, are rarely assessed in current benchmarks. Third, evaluation must attend to temporal structure. Static benchmarks cannot assess how systems behave over extended interactions or maintain consistency as patient data accumulates.
\subsection{Design and Policy Implications}
Our findings suggest several implications for practitioners and policymakers. 

\textit{For design}: constrain systems architecturally rather than by policy alone; preserve legibility of reasoning to support informed oversight; design for workflow integration rather than isolated task performance; and adopt risk-based deployment beginning with low-stakes administrative tasks.

\textit{For policy}: regulatory frameworks should require explicit disclosure of the autonomy level at which systems are designed to operate versus marketed. Systems claiming agentic capabilities should demonstrate those capabilities under conditions representative of actual deployment, including the human oversight structures that will constrain their operation. Liability frameworks should account for the gap between marketed and deployed capabilities, with clearer standards for characterizing autonomy levels to support coherent liability assignment.

\section{Limitations}
This study has several limitations. Our sample of 20 US-based participants, while appropriate for qualitative research seeking thematic saturation~\cite{muller2019data}, represents a specific slice of the healthcare AI ecosystem. Participants were predominantly affiliated with academic medical centers, technology companies, or research institutions. Perspectives from community hospitals, rural settings, safety-net institutions, and international contexts may differ. Our sample also skewed toward individuals with technical backgrounds; frontline clinicians with limited AI exposure were underrepresented.

Certain stakeholder perspectives were less represented, including patients, patient advocates, hospital administrators, and regulators. These groups hold distinct stakes in how agentic AI is defined and deployed. Patient perspectives on autonomy and accountability may diverge significantly from those of clinicians and developers.

Our interviews were conducted between August and December 2025, a period of rapid evolution in agentic AI. The systems and practices participants described may have changed since data collection. Additionally, our findings rely on self-reported experiences rather than direct observation of evaluation practices or clinical workflows. Direct observation of these practices would strengthen future research.

Finally, by directly asking participants to define agentic AI, we may have prompted engagement with a term some do not use in their everyday work. Several participants noted the term felt unfamiliar or artificially imposed which itself is an evidence of the definitional instability we document. However, this also means participants may have engaged more deeply with the concept than they otherwise would. Despite these limitations, our findings provide a grounded account of how agentic AI is conceptualized, evaluated, and constrained across the healthcare AI pipeline.

\section{Conclusion}
Across 20 interviews, we find that `agentic AI' in healthcare is a contested concept. Stakeholders invoke agency differently depending on role and context. This ambiguity shapes which capabilities are prioritized, which failures are anticipated, and how accountability is distributed when systems fall short. At the same time, the systems most often described as agentic are rarely allowed to act autonomously in clinical settings. Oversight, liability, and operational constraints keep these tools in bounded, assistive roles even as public and commercial narratives continue to imply more independence than deployment permits.

This mismatch is sustained by evaluation practices that over-index on task performance while under-assessing the properties that govern safe use in real clinical environments. Stakeholders in the space emphasize the need to measure not only whether outputs are `correct,' but whether systems behave reliably over time, communicate uncertainty, escalate appropriately, and fit into multi-actor workflows where coordination and documentation have downstream consequences. These are precisely the dimensions most likely to determine whether `agentic' systems can be trusted, monitored, and governed in practice.

Taken together, our findings suggest that progress on agentic AI in healthcare hinges on shifting evaluation and governance toward deployment conditions rather than aspirational capability. More precise articulation of autonomy levels, clearer expectations for oversight, and assessments that reflect clinical risk and workflow realities can reduce over-claiming and make accountability legible. Without this shift, the agentic label will continue to promise autonomy while day-to-day systems remain constrained, leaving responsibility ambiguous when things go wrong.

\section*{Generative AI Disclosure Statement}
We used ChatGPT 5.2 to assist with LaTeX formatting, particularly for creating and structuring tables, and for minimal grammatical and stylistic editing.

\begin{acks}
The authors thank Dr. Kameron Black for helpful discussions and guidance, and Daniel Fein for reviewing the manuscript and for his enthusiastic participation during title deliberations. Gabriela Aranguiz Dias is supported through a grant from Accenture.
\end{acks}

\bibliographystyle{ACM-Reference-Format}
\bibliography{AAI}

\appendix
\section{Appendix}
\subsection{Interview Questions}\label{sec:interview_qs}
\begin{table}[H]
\caption{Semi-Structured Interview Protocol}
\label{tab:interview_protocol_final}
\footnotesize
\begin{tabularx}{\columnwidth}{@{} X l @{}}
\toprule
\textbf{Interview Question} & \textbf{Goal / Target} \\ \midrule

\multicolumn{2}{@{} l}{\textbf{Framing and Definitions}} \\
How do you define an ``agentic AI system'' in your domain? & Conceptualizing agency \\
Can you describe any AI systems you currently work with or have recently evaluated? & All Interviewees \\
Do you consider any of these systems to be ``agentic''? Why or why not? & \\
What capabilities or behaviors would make an AI system feel agent-like to you? & \\ \midrule

\multicolumn{2}{@{} l}{\textbf{Technical Capability \& Perceived Autonomy}} \\
To what extent do current AI systems in healthcare operate autonomously vs. under strict human oversight? & Assess limitations \\
What are the current technical bottlenecks in designing more agentic AI systems in healthcare? & Engineers / ML Researchers \\
How do you currently evaluate the effectiveness or safety of these systems---do your metrics account for agent-like behaviors? & \\
Do you foresee shifts from reactive to proactive AI systems in your space? What would drive or limit that shift? & \\ \midrule

\multicolumn{2}{@{} l}{\textbf{Practical Use \& Experience}} \\
Have you worked with or seen AI tools that act independently or proactively? How did that make you feel? & Perception in deployment \\
When an AI system makes an unexpected or autonomous decision, how do you usually respond? & Users / Designers \\
Can you recall a time when you disagreed with an AI's suggestion or action? What happened next? & \\
Do you feel that existing systems adequately explain their goals or actions to you? & \\ \midrule

\multicolumn{2}{@{} l}{\textbf{Trust, Oversight, and Human-AI Relationships}} \\
How do you currently decide when to trust or override an AI system's recommendation? & Trust calibration \\
Do you feel confident that the system shares your goals or understands clinical priorities? & All Interviewees \\
What kinds of transparency or feedback would make you more comfortable with systems acting independently? & \\
How do regulatory, ethical, or institutional constraints shape your willingness to deploy more agentic AI? & \\ \midrule

\multicolumn{2}{@{} l}{\textbf{Gaps, Challenges, and Future Needs}} \\
Where do you see the biggest mismatch between what agentic AI can do vs. what users expect or accept? & Identify agency gaps \\
Are there frameworks, tools, or evaluation methods you feel are missing to bridge this gap? & All Interviewees \\
How do you think trust in agentic AI could be measured? Who should be involved? & \\
What would an ideal agentic AI system look like in your workflow---and what would it take to get there? & \\ \bottomrule
\end{tabularx}
\end{table}
\subsection{Coding Framework}\label{sec:coding_framework}
\begin{table}[H]
\caption{Qualitative Coding Framework (Part I): Definitions, Barriers, and Evaluation}
\label{tab:coding_part1}
\footnotesize
\begin{tabularx}{\columnwidth}{@{} l X @{}}
\toprule
\textbf{Code} & \textbf{Description} \\ \midrule
\textit{01. Definitions and AI Use} & \\
01.agentic\_definition & Definition of agentic AI \\
01.agentic\_characteristics & Characteristics that make a system ``agentic'' \\
01.agent\_capabilities & Capabilities or behaviors that make an AI system feel ``agent-like'' \\
01.agentic.unfamiliar & Interviewee does not know what agentic AI is \\
01.interviewee & Interviewee describes their job/background \\
01.use.ai & Describes experience using AI \\
01.use.ai.agentic & Describes experience using agentic AI \\
01.use.ai.agentic.healthcare & Describes experience using agentic AI in healthcare \\
01.use.ai.positive & Describes a positive view toward an AI tool \\
01.use.ai.negative & Describes a negative view toward an AI tool \\
01.use.ai.correct & Describes how AI tools should be used in healthcare \\ \midrule

\textit{02. Barriers, Evaluation, and Oversight} & \\
02.barriers & Describes a bottleneck to implementing agentic AI in healthcare \\
02.barriers.adoption & Describes and explains adoption potential/barriers \\
02.barriers.technical & Technical bottlenecks to designing/implementing agentic AI in healthcare \\
02.barriers.autonomy & Barriers to the autonomous aspect of agentic AI specifically \\
02.barriers.deployment & Barriers to deploying AI systems in healthcare \\
02.barriers.evaluation & Barriers in creating or performing evaluations \\
02.barriers.transparency & Barriers in providing transparency into agentic AI model workings \\
02.barriers.predicted.integration & Predicted obstacles in integrating agentic AI in healthcare \\
02.evaluation & How to evaluate effectiveness or safety of agentic systems \\
02.evaluation.ideal & Describes what ideal evaluation looks like \\
02.evaluation.metrics & What metrics are used to evaluate deployed AI systems \\
02.evaluation.metrics.technical & Technical evaluation metrics to assess agentic systems \\
02.evaluation.metrics.qualitative & Human-centered or qualitative metrics that assess agentic AI \\
02.evaluation.users & Whether AI users see system metrics and how they feel about them \\
02.evaluations.bad & Evaluations believed to be incorrect, misleading, or poorly performed \\
02.oversight & General description of what oversight is performed, expected, or required \\
02.oversight.strict.human & Human always closely and manually checks AI work for correctness \\
02.oversight.minimal & Situation or use case where oversight is minimal or minimally necessary \\
02.oversight.none & AI systems used without oversight \\
02.oversight.automated & Automated mechanisms determine when to trust or override AI action \\
02.oversight.override & How to decide to trust or override an AI recommendation \\
02.oversight.assumptions & Developer assumptions about oversight level without enforcement \\
02.oversight.depends & Users naturally give more oversight depending on decision impact \\
02.oversight.programaticallybuiltin & Systematic oversight built into agentic systems \\
02.proactive & Possible shifts from reactive to proactive AI \\
02.adoption.pro & Reasons/influences on why agentic AI should/will be adopted \\ \bottomrule
\end{tabularx}
\end{table}

\begin{table}[H]
\caption{Qualitative Coding Framework (Part II): Trust, Concerns, and Future Directions}
\label{tab:coding_part2}
\footnotesize
\begin{tabularx}{\columnwidth}{@{} l X @{}}
\toprule
\textbf{Code} & \textbf{Description} \\ \midrule
\textit{03. Trust and Independence} & \\
03.trust.build & Ways to build more trust \\
03.trust.build.transparency & How transparency could build trust \\
03.trust.build.imposinglimits & Building limitations into AI systems to increase trust \\
03.trust.build.accreditation & Would trust AI more if it used highly accredited information \\
03.trust.measure & How trust in agentic AI could be measured \\
03.trust.measure.who & Who should be involved in evaluating trust \\
03.trust.goals & Do systems share your goals like a human would? \\
03.independence & Experience with AI tools that act independently \\
03.independence.feelings & Feelings in response to independent/proactive AI experience \\
03.independence.build\_comfort & What transparency or feedback would build comfort with independent systems \\
03.explainability & Do existing systems adequately explain their goals or actions? \\ \midrule

\textit{04. Concerns and Constraints} & \\
04.concerns & Concerns about agentic AI being deployed in healthcare \\
04.concerns.serious & Safety-critical concerns (life vs. death) \\
04.concerns.moderate & Concerns that could moderately impact patient care \\
04.concerns.institutional & Institutional concerns about agentic AI deployment \\
04.concerns.integration & Concerns about integrating agentic AI in healthcare \\
04.concerns.useless & Concerns that tools do not save time or introduce more work \\
04.concerns.metrics & Concerns around metric accuracy, use, or influence \\
04.concerns.bias & Concerns AI will create bias toward certain products/services \\
04.concerns.overreliance & Concern that professionals will rely too much on AI and lose skills \\
04.concerns.quality & Concern that using agentic tools will harm work quality \\
04.concerns.training & Concerns that users will not be trained to use AI tools properly \\
04.constraints & How regulatory/ethical/institutional constraints shape willingness to deploy \\ \midrule

\textit{05. Expectation Gaps} & \\
05.expectations.gaps & Gaps between agentic AI capabilities and what users expect or accept \\
05.expectations.align & Frameworks, tools, or evaluation methods to bridge expectation/capability gap \\
05.ideal.agentic & Describes ideal agentic AI in workflow \\ \midrule

\textit{06. Future Directions} & \\
06.goal\_for\_agentic\_ai & Possible outcomes, goals, or vision for agentic AI in healthcare \\
06.expectation\_for\_aai & Expectation for what agentic AI will look like in the future \\
06.development\_incentives & How different incentives drive different stakeholders \\
06.agency\_limitations & Reason or use case for why systems aren't perceived as agentic \\ \midrule

\textit{Utility Codes} & \\
00.potential\_quote & Potential good quote/point \\
00.interesting & Further review required \\ \bottomrule
\end{tabularx}
\end{table}

\end{document}